\documentclass[preprint]{JASA}

\begin{document}

\title[]{Unreported large errors from two PAMGuard three-dimensional localizers of whale calls}


\author{Maya Mathur}
\email{mathur.maya2@gmail.com}
\affiliation{University of Pennsylvania, Philadelphia, PA, 19104, USA}
\author{Luke Stoner-Eby}
\email{lukeseby@seas.upenn.edu}
\affiliation{University of Pennsylvania, Philadelphia, PA, 19104, USA}
\author{Devin Pascoe}
\email{dpascoe@sas.upenn.edu}
\affiliation{Dept. of Earth and Environmental Science, U. of Pennsylvania, Philadelphia, PA 19104, USA}
\author{John L. Spiesberger}
\email{john.spiesberger@gmail.com}
\affiliation{Dept. of Earth and Environmental Science, U. of Pennsylvania, Philadelphia, PA 19104, USA}



\email{john.spiesberger@gmail.com}




\begin{abstract}
Confidence intervals of location (CIL) of calling marine mammals, derived from time-differences-of-arrival (TDOA)
  between receivers, depend on errors
  of TDOAs, receiver location, clocks, sound speeds, and location method. 
  When these errors are minuscule, simulations yield small errors of PAMGuard's 3D simplex localizer when  
  click sounds of beaked and sperm whales  originate in 
  a 1000 x 1000 x 1000 $\mbox{m}^3$  region using five receivers having horizontal and vertical  separations of 1000 m  and 150 m respectively.
  Realistic uncertainties of sound speed up to $\pm 10$ m/s lead to errors up to $10^{14}$ m. 
  With 
  clocks maintained by atomic standards  and common practice of correcting TDOA from synchronization measurements at the start and end of an experiment,
  errors of location
  are up to  $10^{4}$ m. Errors up to $10^2$ and $10^3$ m are found when the receiver's locations are uncertain within 10 and 40 m respectively.
  Errors of PAMGuard's 3D {\it hyperbolic}
  localizer are almost
  independent of the above uncertainties, yielding errors of location up to about $10^4$ m even when simulated errors are minuscule. 
  Causes of PAMGuard's 3D location errors are unknown. 
  These algorithms are briefly compared to another method designed to yield a reliable CIL.
\end{abstract}


\maketitle



\section{\label{sec:1} Introduction}

Locations of sounds can be derived from recordings at widely-separated receivers 
from their TDOA. Many methods of location exist and the accuracy
of location is susceptible to errors in the TDOA, receiver locations, speed of sound, receiver's clocks,
and the location method
itself.  Some methods are used when they are not designed to account for all errors, which may lead to errors in location.
For example,  a method for locating signals via TDOA beamforming yields errors up to 50 km compared with
GPS-located sources \citep{BOEM2018}.  A recent analysis explained the problems,
all originating from the fact 
the method was not designed to account for realistic errors often encountered in marine bioacoustics
work \citep{cse_eval}. The biggest  problem originated from the receiver's unsynchronized clocks.
Another method, commonly referred to as Ishmael, outputs locations with large errors as well,  even when inputs
have minuscule errors \cite{ishmael}.
In this paper, a similar study is made of 3D locations derived from two algorithms distributed via
the PAMGuard web site \cite{pamguard}.
All potential errors are explored here so past and future use may be properly assessed. Another method, called sequential
bound estimation and its related technologies (SBE),
is designed to account for all types of error and is briefly compared with PAMGuard
\citep{prob_distr, sbe, 2d_black_holes}.

The paper is organized as follows.  Sec. \ref{sec:tdoas} summarizes how locations are derived from TDOA in general and specifically for
methods employed by PAMGuard.  Our simulator's design is described in Sec. \ref{sec:simulator}. Sec.
\ref{sec:what_evaluating} provides an overview of how PAMGuard is evaluated. Results of the accuracy of PAMGuard via simulation are presented in
Sec. \ref{sec:simulations}. A discussion appears in Sec. \ref{sec:discussion} and is followed by our conclusions (Sec. \ref{sec:conclusions}).

\section{\label{sec:tdoas} Locations via TDOA}

A TDOA, $\tau$, can be transformed into the difference of distance, $\delta d$, between the calling animal and two receivers using
$ \delta d = c \tau$, when $c$ is a known speed of sound. The locus of points in space sharing this difference in distance 
for a pair of receivers defines a hyperboloid in 3D space \citep{merriam_webster}. Each additional receiver
introduces a new TDOA and its associated hyperboloid.
\textcite{tyrell} notes five receivers are sometimes needed to obtain a mathematically unique solution for location via the intersections
of the hyperboloids, and \textcite{schmidt}
describes the regions in space where five are  needed instead of four.

\subsection{\label{sec:simplex} Simplex method}

The PAMGuard website describes the ``simplex method'' for locating sounds in 3D \cite{pamguard_simplex}. A log-likelihood method for
location is derived accounting for errors in the measured TDOA, receiver locations, and speed of sound
\cite{gillespie_2020}. Errors of the measured TDOA are increased due to uncertainties of the receiver's clocks, locations,
and speed of sound,
and then the log-likelihood of measured and modeled errors is maximized with the simplex method. A correct solution
requires finding a global instead of a local maximum. This is attempted by choosing four different starting points for the
search; the first at the center of the
receiving array, and  the others offset by a random distance in each spatial dimension.  The distance is drawn from a Gaussian distribution whose
width equals the array's maximum aperture \cite{gillespie_2020}.
The algorithm was evaluated experimentally for clock errors, hydrophone locations, and sound speeds varying from  1 to 10 $\mu$s, 0.001  to 0.04 m, and 10 m/s respectively
(Table 1, \citealt{gillespie_2020}).
Locations are derived from TDOA
assuming a one-to-one correspondence between each TDOA and a
hyperboloid \cite{macaulay_2017}. The  speed of sound on each section is the same.
They experimentally  validated their locations with twelve receivers and a
sound source within 50 m  of the hydrophones.    The sound source was lowered from a ship  and the location of the top of the cable derived with
differential GPS.  They reported location errors of 1 to 2 m  horizontally and 5 m vertically \cite{gillespie_2020}.

The inputs to PAMGuard are the receiver locations, a speed of sound, uncertainty of receiver locations, uncertainty for the speed of sound,
and the time  series from each receiver. PAMGuard derives the TDOA from the time series.

\subsection{\label{sec:hyperbolic} Hyperbolic method}

The PAMGuard website \cite{pamguard_hyperbolic} states 3D locations can be obtained using the ``hyperbolic method'' based on 
an analytical solution published by \citet{pass_loc}.  This method is equivalent to intersecting hyperboloids.  The inputs for the
PAMGuard implementation 
are the receiver locations, the speed of sound, and the time series from each receiver. PAMGuard derives the TDOA from the time series.

\section{\label{sec:simulator} Simulator}

\begin{figure}[ht]
  \centerline{\includegraphics[width=6in]{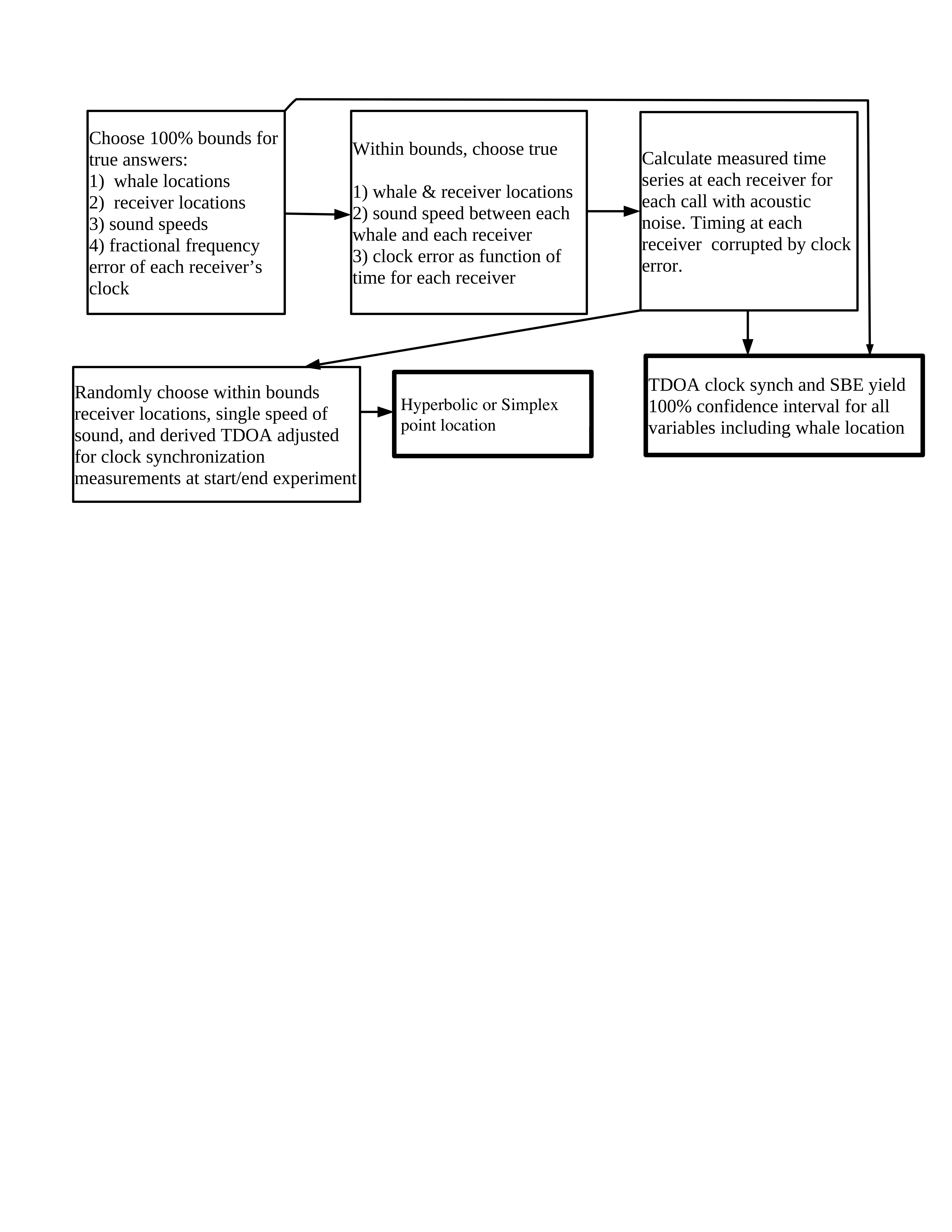}}
\caption{\label{fig:flow_diagram}Flow diagram of simulator. “TDOA clock synch” is \citep{patent_clock_sync1, patent_clock_sync2} and SBE is sequential bound estimation \cite{sbe}.}
\end{figure}

To quantify errors with TDOA-based methods, including those occurring with PAMGuard, a data simulator is constructed (Fig. \ref{fig:flow_diagram}).
The true values of the whale and receiver locations are chosen, as well as the true speeds of sound between the whale and each receiver.
These are used to derive a perfect time series of pressure fluctuations at each receiver, and these are corrupted to simulate actual measurements.
The choice of these true values are chosen within 100\% confidence intervals (CI) determined by the scenario considered. For example, the
CI for receiver location are smaller for a situation where their locations are measured more accurately. Within the CI, the true values are chosen 
following a uniform distribution. The simulator also generates clock errors with slow temporal variation limited by its maximum fractional frequency 
error. Each simulated call is centered at 30 kHz and is modulated with a Gaussian envelope, $\exp(-(t/\tau)^2)$, where $t$ is time and $\tau = 0.2$ ms,
similar to the duration of some click sounds from beaked and sperm whales \cite{zimmer_clicks,beaked_echolocation,sperm_duration}.

Unlike the highly directional click sounds created by beaked and sperm whales \cite{zimmer_clicks,hildebrand_2015,sperm_direction},
the simulator unrealistically provides more data useful for evaluating PAMGuard's location by assuming emitted sounds are omnidirectional.
Thus all  receivers
detect loud click sounds regardless of the swimming direction of a whale.
For each whale call 
and each receiver, an acoustic time series is simulated for perfect clocks and white noise is added, yielding 
a signal-to-noise ratio (SNR) from 30 to 55 dB (not shown).
These time series are provided for input to PAMGuard when the simulation scenario has perfect clocks.
When the receiver's clocks have error, the time series samples are modified via interpolation by subtracting a linear trend from their simulated time stamps.
The linear trend is derived from simulated clock synchronization measurements at the  start and end  of the experiment.
The modified timeseries are presented to PAMGuard.

PAMGuard outputs a location at a point with a CI for the simplex method (Sec. \ref{sec:appendix_conf_interval}), and without a CI for the hyperbolic method.
After the simulation is complete, another computer program compares locations derived with PAMGuard with the true 
answers.  For purposes of scientific reproducibility, the mercurial changeset number is 70 for our simulator and the
comparison program.

\section{\label{sec:what_evaluating} What is being Evaluated in PAMGuard}

Evaluation is made here of the PAMGuard software system, including instructions for its use by the individuals
who maintain the software.  

The input to PAMGuard is comprised of two classes of files, the audio files for each simulated call
and a configuration file defining
how the simulated data are processed to derive locations.  Our simulator  generates the audio files in the wav format.
Section \ref{sec:appendix_config_files}
describes some values we thought to be correct for the configuration file based on PAMGuard's documentation. These may have contributed to PAMGuard
not outputting any locations.
Thus, we sent our configuration file to PAMGuard scientists who sent back a configuration file
that did generate locations, and whose values did not appear to be  consistent with the documentation (Sec. \ref{sec:appendix_config_files}).
Nonetheless,
we used the configuration file yielding locations, with one modification to use the more reliable ``waveform envelope'' method for detecting peaks
in cross-correlations functions \cite{tdoa_gillespie_2019}.

PAMGuard outputs a database containing locations and information about how to associate each location to each input audio file. This
association step is done by finding the nearest time stamp in a database with the time stamp read from the name of each audio
file (Sec. \ref{sec:appendix_association}). Occasionally, there is more than one database time stamp corresponding to a single audio file, and
then we choose the location
nearest the true simulated location.  Locations were then compared to our simulator's true locations.

Early in our evaluation, we were informed the location modules themselves were very difficult to separate from
the software that calls them, and were strongly dissuaded from the attempt.
This prevents a separate evaluation of the
location modules themselves, but has the benefit of evaluating the entire system needed to provide locations.
We could  not find another evaluation of PAMGuard's entire software suite in the literature via simulation.

Unless otherwise noted, we used PAMGuard Version 2.02.28, with computations made with the Ubuntu linux operating system.

\section{\label{sec:simulations} Simulations} 
\begin{figure}[ht]
\centerline{\includegraphics[width=4in]{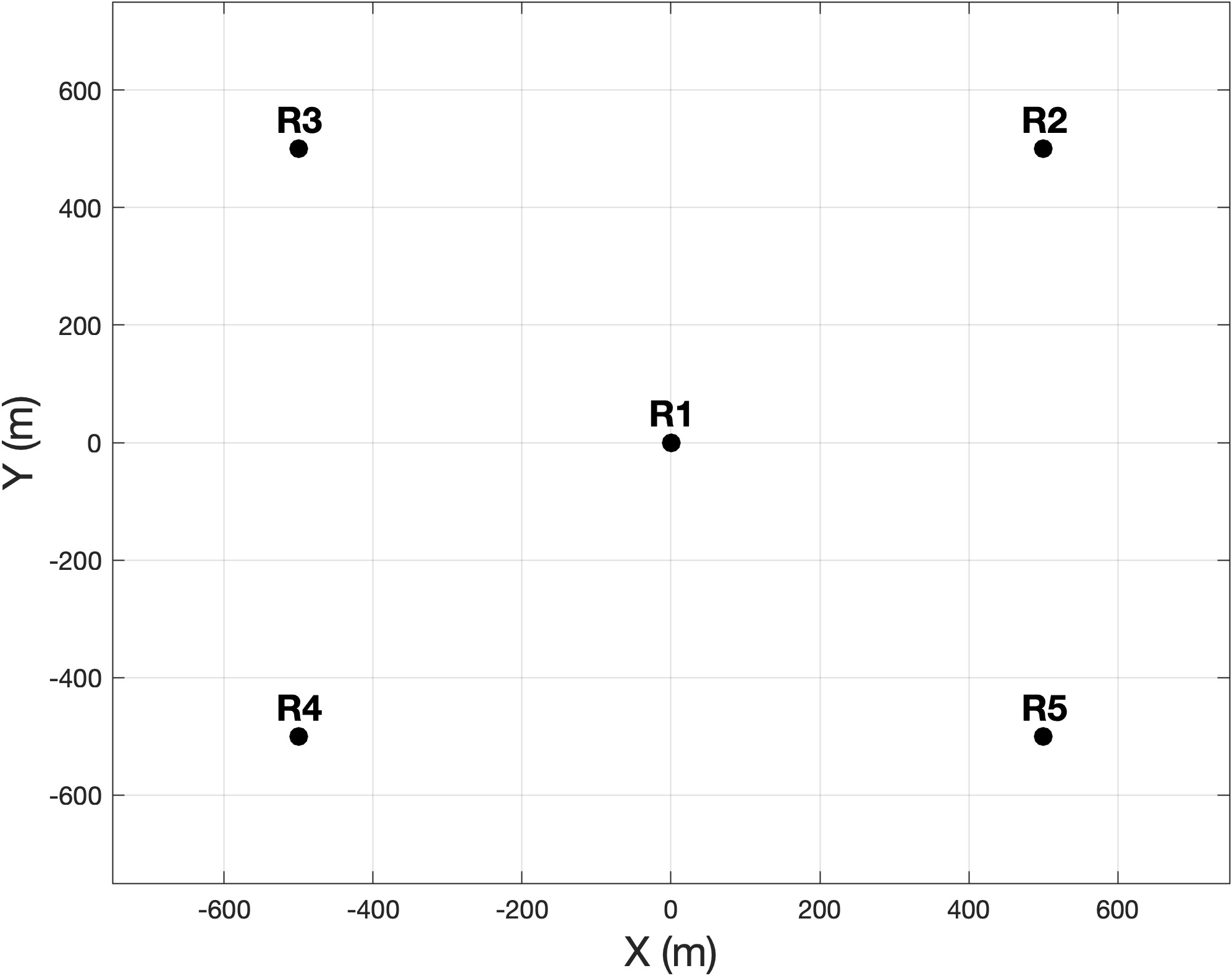}}
\caption{\label{fig:rec_locs_figure} Horizontal coordinates of five receivers. Vertical coordinates in text.}
\end{figure}

Unless otherwise noted, locations of 100 simulated click calls are chosen randomly within a
1 km x 1 km x 1 km volume over a duration of one day, when errors of clocks are minuscule, and 6 mo, when errors of clocks are not minuscule.
The distances within the volume are less than the normal detection distances of beaked and  sperm whales if their beams are
pointing toward a receiver \cite{detection_range,sperm_range}.  Simulated time series are derived for five receivers except one case where four
receivers are simulated
(Fig. \ref{fig:rec_locs_figure}). Their (x,y,z) Cartesian coordinates are: R1: (0, 0, -1000); R2: (500, 500, -900); R3: (-500, 500, -850);
R4: (-500, -500, -800); R5: (500, -500, -970). The vertical axis, $z$, is positive up, and zero at the surface of the ocean.  The PAMGuard simplex method outputs
a  cylinder representing its CIL. The PAMGuard website does not explain or cite a derivation for this CIL. The cylinder's  axis is parallel to the $z$ axis.
We expanded the cylinder's size to correspond to a CIL of 99.5\%.
Error of this CIL is the least distance between the whale and perimeter of the expanded cylinder.
When the whale's true point location is on or inside the expanded cylinder, 
the error of the CIL is zero.  
Errors are displayed with a logarithmic scale with errors equal to 0  underplayed because their logarithm is 
negative infinity.

Error of the PAMGuard hyperbolic method is the distance between the true location and the point location outputted by PAMGuard. PAMGuard does not output a CIL for this method.

\begin{figure}[ht]
\centerline{\includegraphics[width=7in]{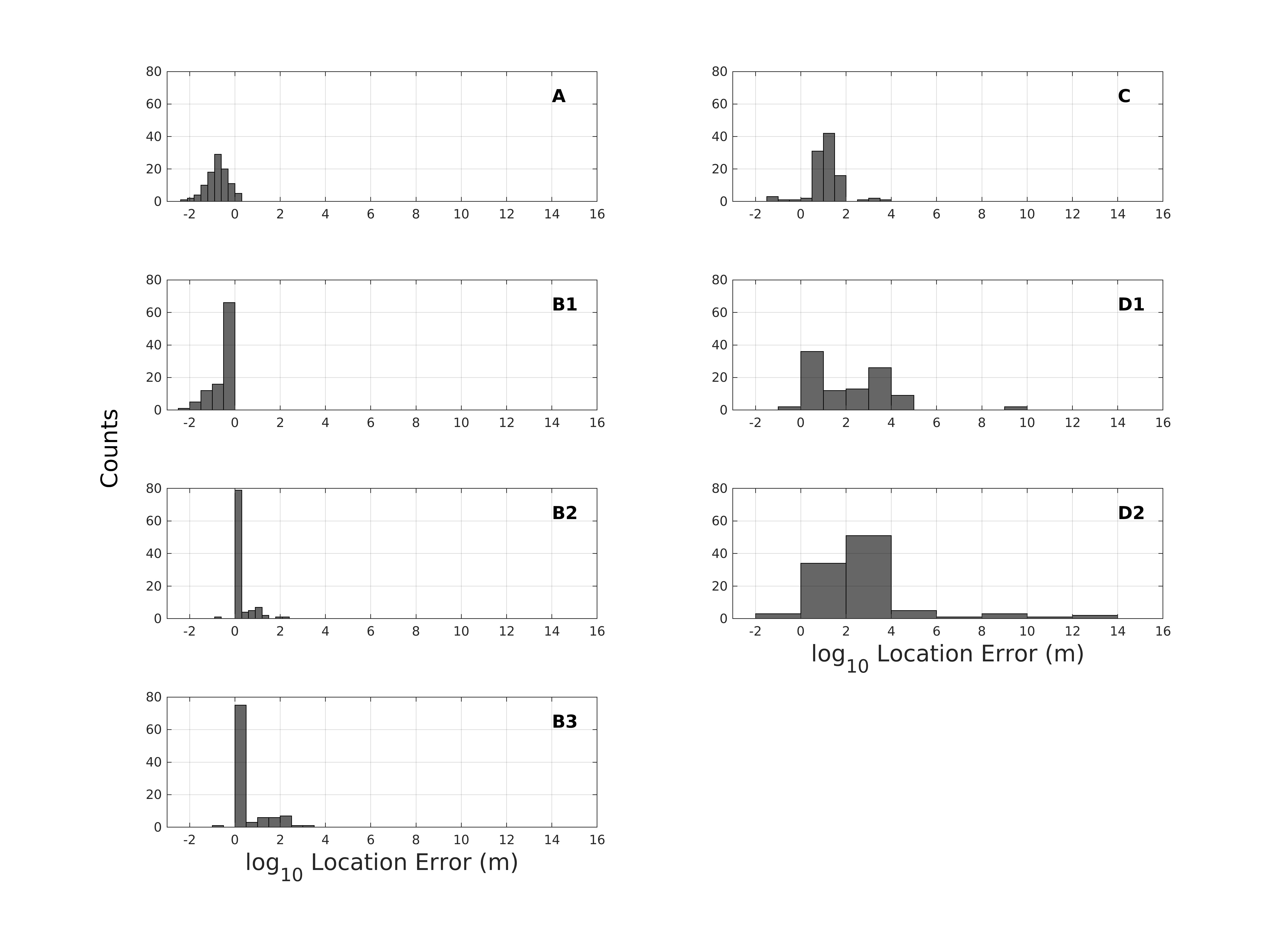}}
\caption{\label{fig:simplex} Location error for 100 simulated whale calls detected on five receivers derived from PAMGuard simplex method for seven error scenarios.
  X axis is $\log_{10}$ of location error in meters; e.g.
  0  and 10 correspond to $10^0$ and $10^{10}$ m respectively. Y axis
is number of counts in each bar. {\bf A}: Minuscule errors for all variables.  {\bf B1-B3}: Uncertainty in receiver position,  $\pm$ 0.05 m,  $\pm$
5 m,  $\pm$ 20 m respectively. {\bf C}: Uncertainty in clock timing $\sim \pm 0.004$ s. {\bf D1-D2}: Uncertainty in speed of sound  of $\pm$ 2.25 m/s and $\pm $ 10 m/s respectively.}
\end{figure}

\begin{figure}[ht]
\centerline{\includegraphics[width=7in]{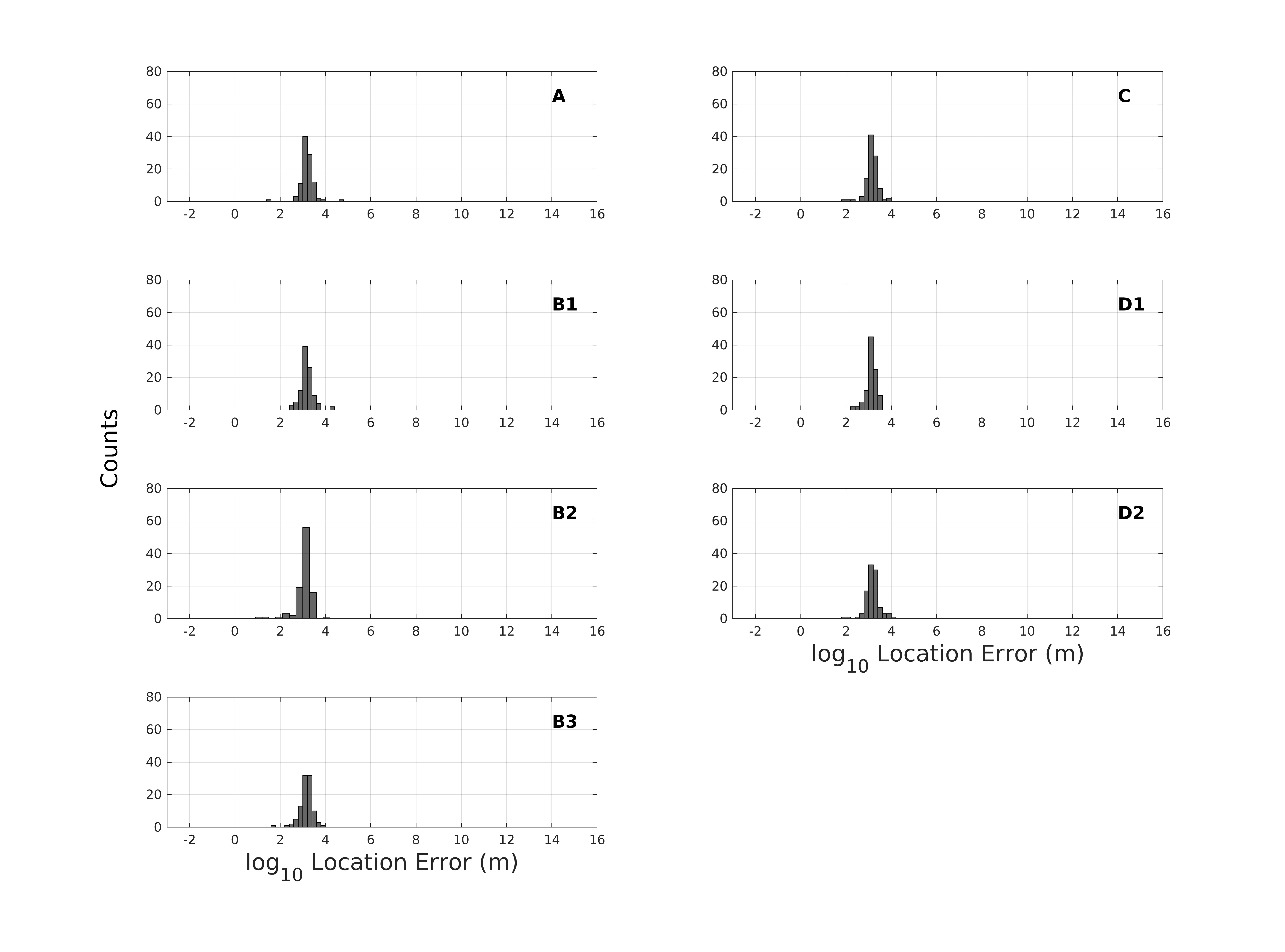}}
\caption{\label{fig:hyperbolic} Same as Fig. \ref{fig:simplex} except for PAMGuard 3D hyperbolic method.}
\end{figure}

An overview of PAMGuard's reliability is provided for each scenario first, with errors rounded to orders of magnitude (Sec. \ref{sec:A}). All scenarios in this section
utilize data from five receivers.
Detailed explanations follow the overview
(Secs. \ref{sec:B}-\ref{sec:E}).

\subsection{\label{sec:A} Overview of PAMGuard's Reliability}

The simplex and hyperbolic methods produce a location for each of the 100 simulated calls for each scenario in Sec. \ref{sec:simulations}.
For scenarios with minuscule errors, the simplex and hyperbolic methods yield errors up to about 1 and $10^5$ m respectively
(Figs. \ref{fig:simplex}A and \ref{fig:hyperbolic}A). 
For receiver position errors up to $\pm 0.05$ m, locations are in error up to  1 m and $10^4$ m  for the simplex and hyperbolic methods respectively
(Figs. \ref{fig:simplex}B1 and \ref{fig:hyperbolic}B1). When  these errors are within $\pm 5$ and $\pm 20$ m, locations from the simplex method are up to about 100 and 1000 m
respectively ((Figs. \ref{fig:simplex}B2-B3). The corresponding errors from the hyperbolic method are up to about $10^4$ m (Fig. \ref{fig:hyperbolic}B2-B3).
When using atomic clock accuracy with  errors up to about 0.004 s maximum during a 6-mo experiment, errors are up to $10^4$ m for both the simplex and hyperbolic methods
(Figs. \ref{fig:simplex}C and \ref{fig:hyperbolic}C).
When the mean speed of sound has error up to 18 m/s with a source-receiver variation up to $\pm 2.25$ m/s,  locations errors are up to  $10^{10}$ m and $10^{3}$ m for the simplex
and hyperbolic methods respectively (Figs. \ref{fig:simplex}D1 and \ref{fig:hyperbolic}D1). When the variation from source to each receiver increases to
$\pm 10$ m/s, location errors are up to $10^{14}$ m and $10^4$ m for the simplex and hyperbolic methods, respectively (Figs. \ref{fig:simplex}D2 and \ref{fig:hyperbolic}D2). 

\subsection{\label{sec:B} Minuscule Errors}
Simulated data were created with minuscule errors of receiver location, sound speed, and clocks. 
The speed of sound ranged from 1450 m/s to 1450.000001 m/s. 
Receivers were placed between 850 m to 1000 m depth with a position error of $10^{-7}$ m (Fig. \ref{fig:rec_locs_figure}).
Maximum clock error was $2 \times 10^{-13}$ s.
Simulations utilized 5 receivers. 

PAMGuard’s simplex 
CIL contained the true location for 4 out of 100 calls (Fig. \ref{fig:simplex}A). The hyperbolic method never contained the true location (Fig. \ref{fig:hyperbolic}A). This follows
because two double precision numbers have nearly zero probability of being the same.
The largest Location errors derived with the simplex and hyperbolic methods are 1.3 m and $4.1 \times 10^4$  m respectively (Figs. \ref{fig:simplex}A,\ref{fig:hyperbolic}A).

\subsection{\label{sec:C} Receiver Position Errors}
PAMGuard was fed horizontal receiver locations randomly selected within cubes of side
lengths 0.1 m, 10 m, and 40 m for three different scenarios respectively.  All other errors for simulated data were minuscule.
PAMGuard’s simplex CIL contained the true location for 100 out of 100 calls for the 0.1 m case, 75 out of 100 calls for the 10 m case, and and 75 out of 100 calls for the 40 m case (Fig. \ref{fig:simplex}B1-B3). PAMGuard’s hyperbolic method contained the true location for 0 out of 100 calls for all receiver position error cases 
(Fig. \ref{fig:hyperbolic}C1-C3).

For the simplex method, the largest errors of location are 0.72 m, 168 m, and 1046 m respectively for the 0,1, 10, and 40 m box length cases respectively (Fig. \ref{fig:simplex}B1-B3).
The corresponding largest errors for the hyperbolic method are $2.2 \times 10^4$ m, $1.5 \times 10^4$ m, and $7.9 \times 10^3$ m (Fig. \ref{fig:hyperbolic}B1-B3).

\subsection{\label{sec:D} Clock Errors}

In the absence of accurate clocks, marine mammal biologists typically
measure their errors at intervals of three to twelve months, and
correct those measurements of the TDOA using a polynomial curve, often a straight line, fit to these measurement instances.
Clock errors are usually measured before and
after deployment and/or by emitting sound at a known location and computing its distance $d_i$ from each receiver $i$ (e.g. p. 12 of \citealt{BOEM2019}).
In the latter situation, a model is used to estimate
the speed of sound, $c_i$ to receiver $i$, yielding an arrival time of $t_i=d_i/c_i$. Receiver $i$'s clock
measures the same arrival
time, $\tilde{t}_i$, an imperfect measurement. Its error is $e_i \equiv \tilde{t}_i - t_i$, where $e_i$ is positive for a fast clock.
A polynomial is fit to the intermittent measurements of $e_i$ and used to estimate the clock offset at any time.  This  method is expensive
to  implement and is not used as frequently as measuring clock errors on the ship  before and after recovery of the instruments.

In the simulation, the errors of each clock were set to vary slowly with time, and were
constrained to not exceed a maximum fractional frequency error, $ \delta f/f = 4.24 \times 10^{-10}$, an obtainable value for atomic clocks.

\begin{figure}[ht]
\centerline{\includegraphics[width=7in]{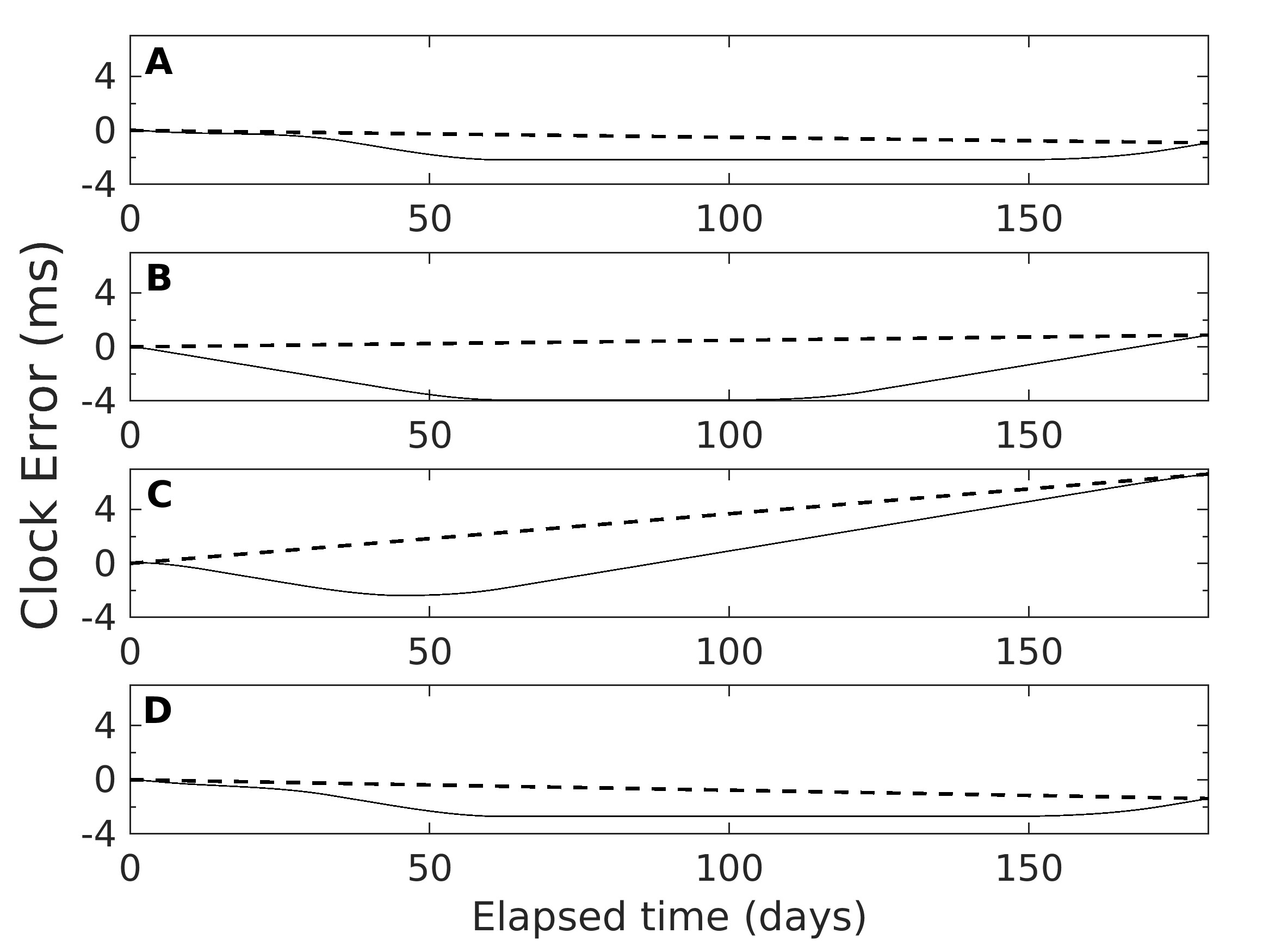}}
\caption{\label{fig:clock_err}Clock error of receiver 1 subtracted from clock errors of receivers 2 through 5 (solid lines) in A–D, respectively. Dashed lines are linear interpolation between synchronization measurements at the beginning and end of experiment. Positive indicates receiver is fast with respect to receiver 1.}
\end{figure}

Time is derived by counting the  number of cycles since some starting time, assuming the oscillator has frequency $f$. The frequency
error of the clock is $\delta f$. It is straightforward to calculate the error of a clock after elapsed time $T$ when $\delta f$ is
a function of time, $t$,  and is $\int_0^T \delta f(t)/f dt$, assuming no error at $t=0$. When $\delta f/f$ is not a function of $t$,
its error at $t=T$ is $T \delta f/f$.
Therefore, for a six-month experiment, clocks can drift up to a maximum of
$T= \pm 6 \times 30 \times 86400 \mbox{s} \times 4.24 \times 10^{-10} = \pm 0.0066$ s. Thus, the maximum fractional frequency error
is useful for understanding how much a clock can be in error.
In the simulator, each clock's error is simulated independently 
so they do not all drift the same way. Their fractional frequency errors are allowed to vary slowly with time, without exceeding
$ 4.24 \times 10^{-10}$, and the integral  mentioned above is used to compute each clock's error at any instant.
When the time from receiver one's clock is subtracted from the others, their clock-error differences drift slowly with values
up to $0.0065$ s (Fig. \ref{fig:clock_err}).  These exhibit slowly-varying non-linear relationships with time as is characteristic with
measurements with oscillators \cite{barnes_1983}.

PAMGuard documentation states clocks should be  within about 1\% of the time required for sound to travel between receivers \cite{pamguard_clock}.
The shortest distance is from receiver R1 to any other (Fig. \ref{fig:rec_locs_figure}), namely 707 m, so taking a speed of sound of 1450 m/s, the propagation time is about 0.5 s.
One percent if this is 0.005 s, and thus the simulated time errors following correction for a linear trend is at this tolerance (Fig. \ref{fig:clock_err}).

The absolute time of a clock has no effect on the accuracy of locating sounds
with TDOA because it disappears when times are subtracted between receivers. Only the differences in the receiver's clocks matter.
Time series for each receiver were  inputted to PAMGuard after correcting the time stamps of each a/d sample using a traditional approach.
The approach removes
a linear trend from the simulated data, the trend derived
from simulated measurements of clock error at the beginning and end of the six-month experiment.
The linear clock-drift approximation has errors of a few  ms (Fig. \ref{fig:clock_err}).
Location errors derived with PAMGuard are up to $4.7 \times 10^{3}$ m for the simplex method and
$6.8 \times 10^{3}$ m for the hyperbolic method (Figs. \ref{fig:simplex}C and \ref{fig:hyperbolic}C).

Clock  errors are also affected by changes in temperature and  physical shock. These effects are not simulated here but could
further degrade the error of location derived with PAMGuard.

\subsection{\label{sec:E} Sound Speed Errors}
For each simulated whale call, the mean speed of sound was randomly chosen in the interval,  [1450 , 1468] m/s. Next,
two scenarios were considered for assigning the speed along the path from said whale to each receiver: the first with $\pm 2.25$
m/s variation about the mean  and the second with $\pm 10$ m/s about the mean, equivalent to  variations of $\pm 0.5^{\circ}$ C and $\pm 2.25^{\circ}$ C respectively. These variations
were not allowed to be outside the interval, [1460 , 1468] m/s.  This procedure was repeated for each whale call.

Others errors were minuscule. The simplex method's CIL contained the true location for 20 and 16 cases out of 100 for the $\pm 2.25$ and $\pm 10$ m/s scenarios respectively.
The corresponding largest errors were $1.5 \times 10^9$ and $7.1 \times 10^{12}$ m respectively  (Figs. \ref{fig:simplex}D1,D2).

The hyperbolic method never coincided with the true location. The largest errors for the  $\pm 2.25$ and $\pm 10$ m/s scenarios were $ 3.2 \times 10^3$ and $1.2 \times 10^4$ m respectively
(Figs. \ref{fig:hyperbolic}D1,2).

\section{\label{sec:discussion} Discussion} 

\subsection{\label{sec:simplex_vrs_hyperbolic} Simplex versus hyperbolic  locations}

Our goal is to report errors of the implementations of the simplex and hyperbolic methods for locating sounds in 3D.
It is beyond the scope of this paper to understand and debug the intricacies of PAMGuard's source code
responsible for picking peaks from cross-correlation functions, implementing the location method, and associating locations with each inputted PAMGuard audio file.
Thus, we do not understand
the cause of the errors, many of which are large (Figs. \ref{fig:simplex},\ref{fig:hyperbolic}). Instead, we follow the advice from PAMGuard scientists
for configuring PAMGuard and interpreting its outputs (Sec. \ref{sec:what_evaluating}, Appendix).

However, we inquired of PAMGuard scientists of a method to check the peak chosen from PAMGuard's generated cross-correlation functions
to see if the correct peak is chosen for finding the TDOAs between each pair of receivers. If the wrong peak is chosen, this might explain why
PAMGuard's errors are sometimes large.
A PAMGuard scientist remarked this is very difficult for us to check from the software tools supplied to the public, including use of their special Matlab functions reading the optional binary files produced by PAMGuard.

When inputs have minuscule errors, large errors of location are found  for the hyperbolic method but not for the simplex method
(Figs. \ref{fig:simplex},\ref{fig:hyperbolic}). Both methods utilize identical audiofiles.  Their PAMGuard configuration files are identical
except for one check box indicating if the 3D locations are derived
with the simplex or hyperbolic method.  PAMGuard's hyperbolic method uses the analytical solution published by \citet{pass_loc}. Those equations
yield the correct locations when there are no uncertainties or errors in data. These facts indicate a problem in PAMGuard's
implementation of the hyperbolic method.  Further evidence of problems with implementation is the observation
that errors from the hyperbolic method are insensitive to the type of simulated error, whether it be receiver location uncertainty,  clock uncertainty, or sound speed uncertainty (Fig.  \ref{fig:hyperbolic}).

The simplex method outputs a CIL, and the error of these CIL increase with uncertainty of receiver location uncertainty, going from a meter to 1046 m when
receiver's locations are uncertain  within $\pm 0.05$ m, $\pm 5$ m, and $\pm 20$ m
(Fig. \ref{fig:simplex}B1-3). When clocks errors are within the published acceptable guidance i.e. a few ms over a 6-mo interval (Sec. \ref{sec:D}),
errors from simplex's CIL are up to $10^4$ m  (Fig. \ref{fig:simplex}C).
When the section-to-section uncertainty in the speed of sound is within  $\pm 2.5$ and $\pm 10$ m/s, the CIL can be in error by up to $10^{10}$ and $10^{14}$ m (Fig. \ref{fig:simplex}D1,D2).
In some cases, the errors derived from
PAMGuard are much larger
than the distances of detection for beaked and sperm whales \cite{detection_range,sperm_range}.

Some of PAMGuard's location errors are 100 to 1000 m (Figs. \ref{fig:simplex},\ref{fig:hyperbolic}).  Unless detected  by simulation,
they might not be large enough to generate 
suspicion, and subsequently be used to derive other scientific conclusions based on flawed information.

\subsection{\label{four_receivers} Four receivers for 3D locations}

\begin{figure}[ht]
\centerline{\includegraphics[width=7in]{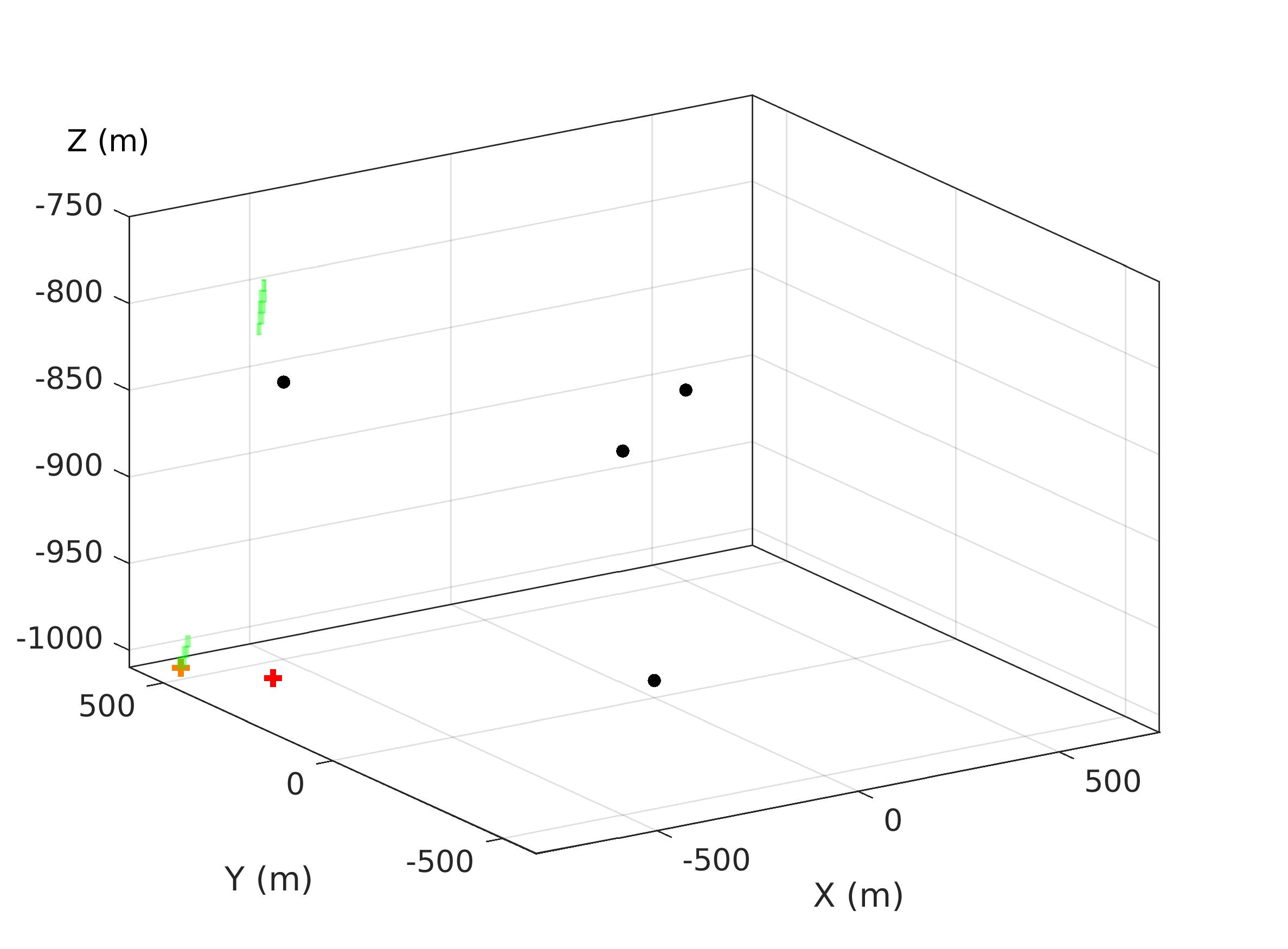}}
\caption{\label{fig:four_rec}Locations of four receivers (black dots), true location of whale (orange) and location derived within PAMGuard 3D simplex method (red).
  100\% CIL derived with SBE (green) with ambiguous locations because only four receivers.}
\end{figure}

With four receivers, there may be two mathematical solutions for location \cite{tyrell,schmidt}. Simulations above utilized PAMGuard version 2.02.18,
but neither the simplex nor hyperbolic methods yielded any locations among the 100 simulated whale calls with four receivers when errors were minuscule..
A previous PAMGuard version, 2.02.11c, did output locations for both methods from four receivers, but never  two locations 
when two mathematical solutions for the equations existed (Fig. \ref{fig:four_rec}). 

In 3D, there are at least three reasons to output all the possible locations when only four receivers are available.   Firstly, no mathematical ambiguity occurs when the call originates
from certain locations \cite{schmidt}.  Secondly, when ambiguous locations occur, one might be excluded because it is above or below the
ocean or it may be too  far away to be detected. Thirdly, if a whale is being tracked, and it enters a region where ambiguous
locations mathematical locations appear, one of these locations might be excluded because it is too far away from the previous
unambiguous location.  

\subsection{\label{isodia_sbe} Location methods designed for all experimental errors via SBE}

Section VB of \citet{ishmael} discusses another approach, known as SBE \cite{prob_distr}, designed to produce reliable 100\% CIL by, in part,
explicitly solving for all  variables affecting location,
including those for clock error (Fig. \ref{fig:flow_diagram}). This reference obviates the  need to repeat further description of its
workings, constituting a completely different approach for locating signals from measurements. This reference explains its 100\% CIL have always been
found to  contain the true location of a source both in many simulations and in an experiment conducted by the US Navy \cite{sbir_phase2}.
For purposes of scientific reproducibility, the following  mercurial changeset implementations of SBE were used for this paper.
Changeset 188 of sbe$\_$2d$\_$analytical and changeset 88 of mstb, where mtsb is the name of one of Spiesberger's software repositories.
All scenarios run with PAMGuard (Sec. \ref{sec:simulations}) were tangentially run with SBE. SBE’s 100\% CIL contained the true simulated location
for all whale calls in each scenario. The development of SBE
was based on the realization there is no point in locating
sounds without reliable confidence intervals.

\section{\label{sec:conclusions} Conclusion}

Past evaluations of the accuracy of the simplex and hyperbolic algorithms demonstrated their reliability for the
scenarios where they were used \cite{gillespie_2020}.  We do not know why the version evaluated here is  unreliable,
yielding errors from 1 to $10^{14}$ m
(Figs. \ref{fig:simplex},\ref{fig:hyperbolic}).
There are many possible explanations for errors occurring in the current software version.
Bugs or other programmatic errors in the software could also have been introduced since it was last evaluated.
\citet{jacobs_2024} used a related algorithm, called Mimplex, to locate sperm whale
clicks within 450 m  of a vertical array.  They state it is a combination of the simplex algorithm and a Markov Chain Monte Carlo
approach.  They did not report checking the Mimplex algorithm's reliability.
Evaluation of PAMGuard's 3D locators here
follows two similar evaluations of the often-used CSE method \cite{cse_eval} and PAMGuard's implementation of  Ishmael 
\cite{ishmael}.  Those also exhibited previously unreported large errors of location.  We recommend
any location algorithm be evaluated with a simulator or experiment prior to
addressing scientific questions, and the implementation version number of the algorithm be published for reproducibility purposes.

An earlier version of PAMGuard, circa July 2023, yielded locations always near the origin of the coordinate system, regardless of the
simulated location of the calling whale (not shown). It is easy to discard use of PAMGuard
software if its locations are obviously incorrect.  The more dangerous scenario is when its outputs cannot easily be determined to be
in error as in
some of the scenarios in Figs. \ref{fig:simplex} and \ref{fig:hyperbolic}. This further motivates employment of a simulator before
deriving locations with any software system.

SBE was designed to account for all errors affecting location, and  was designed to produce extremely reliable CIL \cite{prob_distr}.
In this study, its 100\% CIL  always contained the simulated true location.  

The findings in this paper can be reproduced with the Supplementary Materials discussed in the Appendix.

\section{\label{sec:supp_mat} Supplementary Material}
See supplementary material at \cite{spiesberger_2026_22811500}.



\begin{acknowledgments}
We thank James Gillespie and Jamie Macaulay for their explanations on how to use PAMGuard for the 3D location algorithms investigated in this paper
We thank Tina Yack for providing many hours of tutorials on PAMGuard's use.
Research  was supported by the Office of Naval Research grant N00014-23-1-2336.
\end{acknowledgments}

\section{\label{declarations}Author Declarations}
\subsection{Conflict of Interest}
Sequential bound estimation and its related technologies is a commercial location service. There are no other conflicts of interest.
The data that support the findings of this study are available within the article and its Supplementary material.



\appendix*
\section{Reproducing Results from Simulator and  PAMGuard}
\label{sec:appendix}
        
The simulator is provided as well as the input file needed for each of the simulated scenarios in 
(Figs. \ref{fig:simplex},\ref{fig:hyperbolic}). The file documenting its use is
called SuppPub1.tar in the supplementary materials.  The simulator
needs to be run in a linux operating system.  After unpacking the tar file, the first file to read is tar\_file\_contents.pdf.
The output of each simulation can be many GBytes.   The paths to the audio files and output sqlite3 database needs to be modified by the user so inputs and
outputs of PAMGuard are  correct.   Locations from PAMGuard can be read from the sqlite database. The Simplex 3D locator outputs
location uncertainty, and these are obtained from the sqlite database (Sec. \ref{sec:appendix_conf_interval}).


\subsection{Configuration Files}
\label{sec:appendix_config_files}

We created a PAMGuard configuration file, simplex\_run1\_no\_locs.psfx (supplementary materials) yielding no locations from PAMGuard
for scenario in Fig. \ref{fig:simplex}A.  After sending it to a PAMGuard scientist for advice,
he supplied a configuration file
for use in this paper, after he verified it worked. It is
files\_for\_scenarios/simplex/run\_1/simplex\_run\_1.psfx, contained in the supplementary materials.  We used it except TDOA
were selected from peaks in the cross-correlation function using
PAMGuard's ``waveform envelope''
approach.
This method  maximizes the probability
of choosing the largest peak in the cross-correlation  function \cite{tdoa_gillespie_2019}. For simulated waveforms similar to those used here,
the correct peak appears to be chosen 100\% of the  time (Fig. 3, \citet{tdoa_gillespie_2019}).   The configuration file yielding
no locations is provided for future inquiry of its failure.

Two of the parameters selected by the PAMGuard scientist do not agree with
PAMGuard's documentation and another seems mysterious (Table \ref{tab:config_file}).

The parameter, min click separation, is defined to equal ``either the maximum separation of your hydrophones in samples, or the length of a typical sound from species of interest, or the maximum of both'' \cite{pamguard_click_detector}. The maximum distance between the receivers is 1428.3 m and the speed of sound is about
  1450 m/s.  The audiofiles supplied to PAMGuard have a sample frequency of 160,000 Hz. On top of this, we allowed up to $\pm 0.1$ s of clock  synchronization
  error, so for two receivers this is equivalent to 2(0.1 s $\times$ 160000) = 32000 samples. Thus we get a  min click separation
  equal to 1428.3/1450*160000 + 32000 = 189605.  To be even safer, we used 190620 (Table \ref{tab:config_file}).  The PAMGuard scientist choose 100, thus the value of 100 was used for this paper.

  The parameter, ``Max click length''  is ``set such that click clips are limited to a maximum number of samples.
  Set this so that it is at least greater than the length of a typical click for your species of interest + the separation between
  hydrophones + the presample + the postsample'' \cite{pamguard_click_detector}. The click length in our paper is 0.001 s, equivalent to 160 samples, and the
  pre and post sample values are 160/2=80, so we computed max click length to be 160 + 190620+2/160/2=190940. The PAMGuard scientist used 1024, the value used
  for the simulations in our paper.

  The PAMGuard scientist set the threshold for detecting a click sound to be 10 dB, the value used in our paper (Table \ref{tab:config_file}). We explained to
  the PAMGuard scientists the minimum SNR among all receivers and all whale calls was 29 dB.  We suggested to PAMGuard a threshold of 20 dB as it is very unlikely
  a noise spike could be 20 dB above the noise level.  The PAMGuard scientists nonetheless said we should use 10 dB, and we did.
  For Gaussian noise, noise spikes up to 10 dB
  are common.

  Because the PAMGuard supplied configuration file disagreed with the documentation, we did not try and understand this issue further.
  The reason is as follows.  If the documentation is incorrect, one is left with either
  studying the source code or empirically attempting other sets of parameters.  No attempt was made to understand PAMGuard's source code as it is a lengthy and difficult
  procedure. As for empirical  parameter checks, there are  at least ten parameters in the click detector, and even if each  had only two possible values, there would be
  $2^{10}$ possible
  configuration files to create to see if PAMGuard yielded locations.  Of course, the number of possible values for many
  parameters exceeds two, so it is impossible to empirically try
  all  possible values of the click detector.

\begin{table}[ht]
  \caption{\label{tab:config_file} Values for Click detection parameters derived from documentation \cite{pamguard_click_detector} and PAMGuard scientist.}
  \begin{tabular}{ccc}
    \hline    \hline
Parameter  Name&Authors&PAMGuard\\
&&Scientist\\
\hline
min click separation (samples)&190620&100\\ 
max click length (samples)&190940&1024\\ 
Trigger threshold (dB)&20&10\\
\hline \hline
    \end{tabular}
\end{table}


\subsection{Association Step}
\label{sec:appendix_association}

PAMGuard's outputted database contains a sound acquisition table and a 3D localiser table.  Both contain a coordinated universal time (UTC) column.  When
they differ by less than 10 s, they are declared by our software to be a correct association, and then the UTC value
is linked to the PAMGuard wav file by finding the PAMGuard wav file name whose embedded time stamp is close to the UTC value.
In a few cases, there are a plurality of UTC values from the sound acquisition table within 10 s of a UTC time from
the 3D localiser table. 
In this case, the location from the localiser table is chosen to be the one nearest the
true location of the simulated whale call.  Our simulator avoids time stamp ambiguity by creating pamguard wav files with embedded time stamps about 17 min apart.

Locations outputted by PAMGuard are in the supplementary materials file called compare\_pamguard\_sbe.mat.  See public\_view\_files\_for\_scenarios.pdf
for its interpretation.  The association between simulated whale call  number, PAMGuard wav file name, and PAMGuard location can be derived from this
matlab file.

\subsection{Confidence Interval of Location with Simplex Method}
\label{sec:appendix_conf_interval}

The database from PAMGuard's includes a one standard deviation of location error represented by a vertical cylinder of radius $r$ and height $h$.
These dimensions are derived from the
Group\_3D\_Localiser table with column heading ``Error1''.  This column contains a variable called ``ANGLES'' with three entries called a(1) through a(3). It also contain a variable called  ``ERRORS'' with six entries, called
e(1) through e(6) here . First, form er=max(e(1) e(4)). Then $r=\cos(a(2))*er$ and $h=\sin(a(2))*er$. PAMGuard errors in this paper are three standard deviation, a multiple of this one standard deviation cylinder.








\bibliography{sampbib}

\begin{thebibliography}{61}
\def\enquote#1{``#1,''}
\def\plainquote#1{``#1''}
\expandafter\ifx\csname natexlab\endcsname\relax\def\natexlab#1{#1}\fi
\providecommand{\dourl}[1]{\href{http://#1}{\nolinkurl{#1}}}
\providecommand{\bibinfo}[2]{#2}
\providecommand{\noopsort}[1]{}
\providecommand{\switchargs}[2]{#2#1}
  \def\eatspace #1{#1}

\bibitem[{Bailey \emph{et~al.}(12)Bailey, Fandel, Silva, Gryzb, McDonald,
  Hoover, Ogburn, and Rice}]{bailey_rice}
\bibinfo{author}{Bailey, H.}, \bibinfo{author}{Fandel, A.~D.},
  \bibinfo{author}{Silva, K.}, \bibinfo{author}{Gryzb, E.},
  \bibinfo{author}{McDonald, E.}, \bibinfo{author}{Hoover, A.~L.},
  \bibinfo{author}{Ogburn, M.~B.},  and \bibinfo{author}{Rice, A.~N.}
  (\textbf{\bibinfo{year}{12}}). \enquote{\bibinfo{title}{Identifying and
  predicting occurrence and abundance of a vocal animal species based on
  individually specific calls}} \bibinfo{journal}{Ecosphere} \textbf{8}.

\bibitem[{Bailey \emph{et~al.}(2019)Bailey, Rice, Wingfield, Hodge, Estabrook,
  Hawthorne, Garrod, Fandel, Fouda, McDonald, Grzyb, Fletcher, and
  Hoover}]{BOEM2018}
\bibinfo{author}{Bailey, H.}, \bibinfo{author}{Rice, A.},
  \bibinfo{author}{Wingfield, J.~E.}, \bibinfo{author}{Hodge, K.~B.},
  \bibinfo{author}{Estabrook, B.~J.}, \bibinfo{author}{Hawthorne, D.},
  \bibinfo{author}{Garrod, A.}, \bibinfo{author}{Fandel, A.~D.},
  \bibinfo{author}{Fouda, L.}, \bibinfo{author}{McDonald, E.},
  \bibinfo{author}{Grzyb, E.}, \bibinfo{author}{Fletcher, W.},  and
  \bibinfo{author}{Hoover, A.~L.} (\textbf{\bibinfo{year}{2019}}).
  \enquote{\bibinfo{title}{Determining habitat use by marine mammals and
  ambient noise levels using passive acoustic monitoring offshore of maryland}}
  \bibinfo{type}{OCS Study BOEM 2019-018}.

\bibitem[{Barlow and Griffiths(2017)}]{detection_range}
\bibinfo{author}{Barlow, J.},  and \bibinfo{author}{Griffiths, E.~T.}
  (\textbf{\bibinfo{year}{2017}}). \enquote{\bibinfo{title}{Precision and bias
  in estimating detection distances for beaked whale echolocation clicks using
  a two-element vertical hydrophone array}} \bibinfo{journal}{Journal of the
  Acoustical Society of America} \dodoi{10.1121/1.4985109}.

\bibitem[{Barnes(1983)}]{barnes_1983}
\bibinfo{author}{Barnes, J.~A.} (\textbf{\bibinfo{year}{1983}}).
  \enquote{\bibinfo{title}{The measurement of linear frequency drift in
  oscillators}} in \emph{\bibinfo{booktitle}{Proceedings of the 15th Annual
  Precise Time and Time Interval Systems and Applications Meeting}}, pp.
  \bibinfo{pages}{551--582}.

\bibitem[{Baumann-Pickering~S(2010)}]{beaked_echolocation}
\bibinfo{author}{Baumann-Pickering~S, Wiggins~SM, R. E. R. M. S. H. H.~J.}
  (\textbf{\bibinfo{year}{2010}}). \enquote{\bibinfo{title}{Echolocation
  signals of beaked whales at palmyra atoll}} \bibinfo{journal}{Journal of the
  Acoustical Society of America} .

\bibitem[{Birchfield(2004)}]{conf_4}
\bibinfo{author}{Birchfield, S.~T.} (\textbf{\bibinfo{year}{2004}}).
  \enquote{\bibinfo{title}{A unifying framework for acoustic localization}} in
  \emph{\bibinfo{booktitle}{2004 12th European Signal Processing Conference}},
  pp. \bibinfo{pages}{1127--1130}.

\bibitem[{Birchfield and Gillmor(2002)}]{conf_3}
\bibinfo{author}{Birchfield, S.~T.},  and \bibinfo{author}{Gillmor, D.~K.}
  (\textbf{\bibinfo{year}{2002}}). \enquote{\bibinfo{title}{Fast bayesian
  acoustic localization}} in \emph{\bibinfo{booktitle}{2002 IEEE International
  Conference on Acoustics, Speech, and Signal Processing}}, Vol. 2, pp.
  \bibinfo{pages}{II--1793--II--1796}.

\bibitem[{Clark \emph{et~al.}(2023)Clark, Charif, Hawthorne, Rahaman, Givens,
  George, and Muirhead}]{clark_charif}
\bibinfo{author}{Clark, C.~W.}, \bibinfo{author}{Charif, R.~A.},
  \bibinfo{author}{Hawthorne, D.}, \bibinfo{author}{Rahaman, A.},
  \bibinfo{author}{Givens, G.~H.}, \bibinfo{author}{George, J.~C.},  and
  \bibinfo{author}{Muirhead, C.~A.} (\textbf{\bibinfo{year}{2023}}).
  \enquote{\bibinfo{title}{Acoustic data from the spring 2011 bowhead whale
  census at point barrow, alaska}} \bibinfo{journal}{International Whaling
  Commission} \dodoi{10.47536/jcrm.v19i1.413}.

\bibitem[{Clark \emph{et~al.}(2010)Clark, Ellison, Hatch, Merrick, Parijs, and
  Wiley}]{report1}
\bibinfo{author}{Clark, C.~W.}, \bibinfo{author}{Ellison, W.~T.},
  \bibinfo{author}{Hatch, L.~T.}, \bibinfo{author}{Merrick, R.~L.},
  \bibinfo{author}{Parijs, S. M.~V.},  and \bibinfo{author}{Wiley, D.~N.}
  (\textbf{\bibinfo{year}{2010}}). \enquote{\bibinfo{title}{An ocean observing
  system for large-scale monitoring and mapping of noise throughout the
  stellwagen bank national marine sanctuary}} \bibinfo{type}{Technical Report}.

\bibitem[{Collier \emph{et~al.}(2010)Collier, Kirschel, and Taylor}]{collier}
\bibinfo{author}{Collier, T.~C.}, \bibinfo{author}{Kirschel, A.~N.},  and
  \bibinfo{author}{Taylor, C.~E.} (\textbf{\bibinfo{year}{2010}}).
  \enquote{\bibinfo{title}{Acoustic localization of antbirds in a mexican
  rainforest using a wireless sensor network}} \bibinfo{journal}{J. Acoust.
  Soc. Am.} \textbf{128}, \bibinfo{pages}{182--189}.

\bibitem[{Fandel \emph{et~al.}(2022)Fandel, Hodge, Rice, and Bailey}]{conf_2}
\bibinfo{author}{Fandel, A.}, \bibinfo{author}{Hodge, K.},
  \bibinfo{author}{Rice, A.},  and \bibinfo{author}{Bailey, H.}
  (\textbf{\bibinfo{year}{2022}}). \enquote{\bibinfo{title}{Altered spatial
  distribution of a marine top predator under elevated ambient sound
  conditions}} in \emph{\bibinfo{booktitle}{State of the Science Workshop on
  Wildlife and Offshore Wind Energy 2022}}, \bibinfo{organization}{New York
  State Energy Research and Development Authority}.

\bibitem[{Friedman \emph{et~al.}(1981)Friedman, Furberg, DeMets, Reboussin, and
  Granger}]{friedman}
\bibinfo{author}{Friedman, L.~M.}, \bibinfo{author}{Furberg, C.~D.},
  \bibinfo{author}{DeMets, D.~L.}, \bibinfo{author}{Reboussin, D.~M.},  and
  \bibinfo{author}{Granger, C.~B.} (\textbf{\bibinfo{year}{1981}}).
  \emph{\bibinfo{title}{Fundamentals of Clinical Trials}}
  (\bibinfo{publisher}{Springer}).

\bibitem[{Gillepsie \emph{et~al.}(2020)Gillepsie, Palmer, Macaulay, Sparling,
  and Hatie}]{gillespie_2020}
\bibinfo{author}{Gillepsie, D.}, \bibinfo{author}{Palmer, L.},
  \bibinfo{author}{Macaulay, J.}, \bibinfo{author}{Sparling, C.},  and
  \bibinfo{author}{Hatie, G.} (\textbf{\bibinfo{year}{2020}}).
  \enquote{\bibinfo{title}{Passive acoustic methods for tracking the 3{D}
  movements of small cetaceans around marine structures}}
  \bibinfo{journal}{{PL}o{S} {ONE}} \textbf{AES-8},
  \dodoi{10.1371/journal.pone.0229058}.

\bibitem[{Gillespie and Macaulay(2019)}]{tdoa_gillespie_2019}
\bibinfo{author}{Gillespie, D.},  and \bibinfo{author}{Macaulay, J.}
  (\textbf{\bibinfo{year}{2019}}). \enquote{\bibinfo{title}{Time of arrival
  difference estimation for narrow band high frequency echolocation clicks}}
  \bibinfo{journal}{J. Acoust. Soc. of Am.} \textbf{146}(4),
  \bibinfo{pages}{EL387--EL392}, \dourl{https://doi.org/10.1121/1.5129678},
  \dodoi{10.1121/1.5129678}.

\bibitem[{Greene \emph{et~al.}(2016)Greene, Shiu, Morano, Clark, Little,
  Billings, and Hawthrone}]{conf_1}
\bibinfo{author}{Greene, E.}, \bibinfo{author}{Shiu, Y.},
  \bibinfo{author}{Morano, J.}, \bibinfo{author}{Clark, C.},
  \bibinfo{author}{Little, P.}, \bibinfo{author}{Billings, A.},  and
  \bibinfo{author}{Hawthrone, D.} (\textbf{\bibinfo{year}{2016}}).
  \enquote{\bibinfo{title}{A practical guide for designing recording arrays in
  terrestrial environments: Best practices for maximizing location accuracy and
  precision}} in \emph{\bibinfo{booktitle}{Ecoacoustics Congress 2016}},
  \bibinfo{organization}{International Society of Ecoacoustics (ISE)}.

\bibitem[{Hildebrand \emph{et~al.}(2015)Hildebrand, Baumann-Pickering, Frasier,
  Trickey, Merkens, Wiggins, McDonald, Garrison, Harris, Marques, and
  Thomas}]{hildebrand_2015}
\bibinfo{author}{Hildebrand, J.~A.}, \bibinfo{author}{Baumann-Pickering, S.},
  \bibinfo{author}{Frasier, K.~E.}, \bibinfo{author}{Trickey, J.~S.},
  \bibinfo{author}{Merkens, K.~P.}, \bibinfo{author}{Wiggins, S.~M.},
  \bibinfo{author}{McDonald, M.~A.}, \bibinfo{author}{Garrison, L.~P.},
  \bibinfo{author}{Harris, D.}, \bibinfo{author}{Marques, T.~A.},  and
  \bibinfo{author}{Thomas, L.} (\textbf{\bibinfo{year}{2015}}).
  \enquote{\bibinfo{title}{Passive acoustic monitoring of beaked whale
  densities in the gulf of mexico}} \bibinfo{journal}{Scientific Reports}
  \textbf{5}(1), \bibinfo{pages}{16343}, \dodoi{10.1038/srep16343}.

\bibitem[{Hoeck \emph{et~al.}(2023)Hoeck, Rowell, Dean, Rice, and
  Parijs}]{hoeck_rowell}
\bibinfo{author}{Hoeck, R. V.~V.}, \bibinfo{author}{Rowell, T.~J.},
  \bibinfo{author}{Dean, M.~J.}, \bibinfo{author}{Rice, A.~N.},  and
  \bibinfo{author}{Parijs, S. M.~V.} (\textbf{\bibinfo{year}{2023}}).
  \enquote{\bibinfo{title}{Comparing atlantic cod temporal spawning dynamics
  across a biogeographic boundary: Insights from passive acoustic monitoring}}
  \bibinfo{type}{Technical Report}, \dodoi{10.1002/mcf2.10226}.

\bibitem[{Jacobs \emph{et~al.}(2024)Jacobs, Gero, Malinka, T{\o}nnesen,
  Beedholm, DeRuiter, and Madsen}]{jacobs_2024}
\bibinfo{author}{Jacobs, E.~R.}, \bibinfo{author}{Gero, S.},
  \bibinfo{author}{Malinka, C.~E.}, \bibinfo{author}{T{\o}nnesen, P.~H.},
  \bibinfo{author}{Beedholm, K.}, \bibinfo{author}{DeRuiter, S.},  and
  \bibinfo{author}{Madsen, P.~T.} (\textbf{\bibinfo{year}{2024}}).
  \enquote{\bibinfo{title}{The active space of sperm whale codas: inter-click
  information for intra-unit communication}} \bibinfo{journal}{J. Exp. Biol.}
  \textbf{227}, \dodoi{10.1242/jeb.246442}.

\bibitem[{john.spiesberger@gmail.com()}]{sbe_service}
\bibinfo{author}{john.spiesberger@gmail.com}.
  \plainquote{\bibinfo{title}{Scientific innovations, inc.}} .

\bibitem[{Macaulay \emph{et~al.}(2017)Macaulay, Gordon, Gillespie, Malinka, and
  Northridge}]{macaulay_2017}
\bibinfo{author}{Macaulay, J.}, \bibinfo{author}{Gordon, J.},
  \bibinfo{author}{Gillespie, D.}, \bibinfo{author}{Malinka, C.},  and
  \bibinfo{author}{Northridge, S.} (\textbf{\bibinfo{year}{2017}}).
  \enquote{\bibinfo{title}{Passive acoustic methods for fine-scale tracking of
  harbour purposises in tidal rapids}} \bibinfo{journal}{J. Acoust. Soc. Am.}
  \textbf{141}, \bibinfo{pages}{1120--1132}.

\bibitem[{Mathur \emph{et~al.}(2024)Mathur, Spiesberger, and Pascoe}]{cse_eval}
\bibinfo{author}{Mathur, M.}, \bibinfo{author}{Spiesberger, J.},  and
  \bibinfo{author}{Pascoe, D.} (\textbf{\bibinfo{year}{2024}}).
  \enquote{\bibinfo{title}{Confidence intervals of location for marine mammal
  calls via time-differences-of-arrival: Sensitivity analysis}}
  \bibinfo{journal}{JASA Express Lett.} \textbf{4}.

\bibitem[{Mennill \emph{et~al.}(2006)Mennill, Burt, Fristrup, and
  Vehrencamp}]{mennil}
\bibinfo{author}{Mennill, D.~J.}, \bibinfo{author}{Burt, J.~M.},
  \bibinfo{author}{Fristrup, K.~M.},  and \bibinfo{author}{Vehrencamp, S.~L.}
  (\textbf{\bibinfo{year}{2006}}). \enquote{\bibinfo{title}{Accuracy of an
  acoustic location system for monitoring the position of duetting songbirds in
  tropical forest}} \bibinfo{journal}{J. Acoust. Soc. Am} \textbf{119},
  \bibinfo{pages}{283202839}.

\bibitem[{Merriam-Webster(2023)}]{merriam_webster}
\bibinfo{author}{Merriam-Webster} (\textbf{\bibinfo{year}{2023}}).
  \plainquote{\bibinfo{title}{hyperbola}}
  \dourl{https://www.merriam-webster.com/dictionary/hyperbola}.

\bibitem[{Miller and Miller(2018)}]{sperm_range}
\bibinfo{author}{Miller, B.~S.},  and \bibinfo{author}{Miller, E.~J.}
  (\textbf{\bibinfo{year}{2018}}). \enquote{\bibinfo{title}{The seasonal
  occupancy and diel behaviour of antarctic sperm whales revealed by acoustic
  monitoring}} \bibinfo{journal}{Scientific Reports}
  \dodoi{10.1038/s41598-018-23752-1}.

\bibitem[{M{\o}hl~B(2003)}]{sperm_duration}
\bibinfo{author}{M{\o}hl~B, Wahlberg~M, M. P. H. A. L.~A.}
  (\textbf{\bibinfo{year}{2003}}). \enquote{\bibinfo{title}{The monopulsed
  nature of sperm whale clicks}} \bibinfo{journal}{Journal of the Acoustical
  Society of America} \dodoi{10.1121/1.1586258}.

\bibitem[{of~Defense(2011)}]{trl}
\bibinfo{author}{of~Defense, D.} (\textbf{\bibinfo{year}{2011}}).
  \plainquote{\bibinfo{title}{Technology readiness levels}}
  \dourl{https://www.ncbi.nlm.nih.gov/books/NBK201356/}.

\bibitem[{PAMGuard(2024)}]{pamguard}
\bibinfo{author}{PAMGuard} (\textbf{\bibinfo{year}{2024}}).
  \plainquote{\bibinfo{title}{Pamguard}} \dourl{https://www.pamguard.org/}.

\bibitem[{{PAMG}uard(2025{\natexlab{a}})}]{pamguard_hyperbolic}
\bibinfo{author}{{PAMG}uard} (\textbf{\bibinfo{year}{2025}}{\natexlab{a}}).
  \plainquote{\bibinfo{title}{Hyperbolic}}
  \dourl{https://www.pamguard.org/olhelp/localisation/group3d/docs/3dhyperbolic.html}.

\bibitem[{{PAMG}uard(2025{\natexlab{b}})}]{pamguard_simplex}
\bibinfo{author}{{PAMG}uard} (\textbf{\bibinfo{year}{2025}}{\natexlab{b}}).
  \plainquote{\bibinfo{title}{Simplex}}
  \dourl{https://www.pamguard.org/olhelp/localisation/group3d/docs/3dsimplex.html}.

\bibitem[{PAMGuard(2026{\natexlab{a}})}]{pamguard_clock}
\bibinfo{author}{PAMGuard} (\textbf{\bibinfo{year}{2026}}{\natexlab{a}}).
  \plainquote{\bibinfo{title}{overview}}
  \dourl{https://www.pamguard.org/olhelp/localisation/group3d/docs/3doverview.html}.

\bibitem[{PAMGuard(2026{\natexlab{b}})}]{pamguard_click_detector}
\bibinfo{author}{PAMGuard} (\textbf{\bibinfo{year}{2026}}{\natexlab{b}}).
  \plainquote{\bibinfo{title}{Pamguard click detector overview}}
  \dourl{https://www.pamguard.org/olhelp/detectors/clickDetectorHelp/docs/ClickDetector_clickDetector.html}.

\bibitem[{Pascoe \emph{et~al.}(2024)Pascoe, Spiesberger, and
  Mellinger}]{ishmael}
\bibinfo{author}{Pascoe, D.}, \bibinfo{author}{Spiesberger, J.},  and
  \bibinfo{author}{Mellinger, D.} (\textbf{\bibinfo{year}{2024}}).
  \enquote{\bibinfo{title}{Unreported large errors in a common method for
  soundd source localization of marine mammals}} \bibinfo{journal}{J. Acoust.
  Soc. Am} \textbf{156}, \bibinfo{pages}{3780--3787},
  \dodoi{10.1121/10.0034547}.

\bibitem[{Review(2018)}]{temp_var}
\bibinfo{author}{Review, T. I.~H.} (\textbf{\bibinfo{year}{2018}}).
  \enquote{\bibinfo{title}{Determination of velocity of sound in seawater in
  cape cod bay}} \bibinfo{journal}{The International Hydrographic Review} .

\bibitem[{Risch \emph{et~al.}(2014)Risch, Siebert, and Parijs}]{risch_siebert}
\bibinfo{author}{Risch, D.}, \bibinfo{author}{Siebert, U.},  and
  \bibinfo{author}{Parijs, S. M.~V.} (\textbf{\bibinfo{year}{2014}}).
  \enquote{\bibinfo{title}{Individual calling behaviour and movements of north
  atlantic minke whales (balaenoptera acutorostrata)}} \bibinfo{journal}{Brill}
  \textbf{151}.

\bibitem[{Salisbury \emph{et~al.}(2018)Salisbury, Estabrook, Klinck, and
  Rice}]{BOEM2019}
\bibinfo{author}{Salisbury, D.~P.}, \bibinfo{author}{Estabrook, B.~J.},
  \bibinfo{author}{Klinck, H.},  and \bibinfo{author}{Rice, A.~N.}
  (\textbf{\bibinfo{year}{2018}}). \enquote{\bibinfo{title}{Understanding
  marine mammal presence in the virginia offshore wind energy area}}
  \bibinfo{type}{OCS Study BOEM 2019-007}.

\bibitem[{Schmidt(1972)}]{schmidt}
\bibinfo{author}{Schmidt, R.~O.} (\textbf{\bibinfo{year}{1972}}).
  \enquote{\bibinfo{title}{A new approach to geometry of range difference
  location}} \bibinfo{journal}{IEEE Trans. on Aerospace and Elect. Sys.}
  \textbf{AES-8}, \bibinfo{pages}{821--835}.

\bibitem[{Spiesberger(15)}]{patent_sbe1}
\bibinfo{author}{Spiesberger, J.} (\textbf{\bibinfo{year}{15}}).
  \plainquote{\bibinfo{title}{Estimation algorithms and location techniques}}
  \bibinfo{howpublished}{U.S. patent 7,219,032}.

\bibitem[{Spiesberger(2004)}]{isodiachrons}
\bibinfo{author}{Spiesberger, J.} (\textbf{\bibinfo{year}{2004}}).
  \enquote{\bibinfo{title}{Geometry of locating sounds from differences in
  travel time: isodiachrons}} \bibinfo{journal}{J. Acoust. Soc. Am}
  \textbf{116}(5), \bibinfo{pages}{3168--3167}.

\bibitem[{Spiesberger(2005{\natexlab{a}})}]{prob_distr}
\bibinfo{author}{Spiesberger, J.}
  (\textbf{\bibinfo{year}{2005}}{\natexlab{a}}).
  \enquote{\bibinfo{title}{Probability distributions for locations of calling
  animals, receivers, sound speeds, winds, and data from travel time
  differences}} \bibinfo{journal}{J. Acoust. Soc. Am.} \textbf{118},
  \bibinfo{pages}{1790--1800}.

\bibitem[{Spiesberger(2005{\natexlab{b}})}]{sbe}
\bibinfo{author}{Spiesberger, J.}
  (\textbf{\bibinfo{year}{2005}}{\natexlab{b}}).
  \enquote{\bibinfo{title}{Probability distributions for locations of calling
  animals, receivers, sound speeds, winds, and data from travel time
  differences}} \bibinfo{journal}{J. Acoust. Soc. Am} \textbf{118},
  \bibinfo{pages}{1790--1800}.

\bibitem[{Spiesberger(2020)}]{2d_black_holes}
\bibinfo{author}{Spiesberger, J.} (\textbf{\bibinfo{year}{2020}}).
  \enquote{\bibinfo{title}{Dimension reduction in location estimation - the
  need for variable propagation speed}} \bibinfo{journal}{Acoustical Physics}
  \textbf{66}, \bibinfo{pages}{178--190}.

\bibitem[{Spiesberger(2026)}]{spiesberger_2026_22811500}
\bibinfo{author}{Spiesberger, J.} (\textbf{\bibinfo{year}{2026}}).
  \plainquote{\bibinfo{title}{Data to duplicate outputs of pamguard from
  paper}} \dourl{https://doi.org/10.5281/zenodo.22811500},
  \dodoi{10.5281/zenodo.22811500}.

\bibitem[{Spiesberger \emph{et~al.}(2021)Spiesberger, Berchok, Iyer, Schoeny,
  Sivakumar, Woodrich, Yang, and Zhu}]{pass_acous}
\bibinfo{author}{Spiesberger, J.}, \bibinfo{author}{Berchok, C.},
  \bibinfo{author}{Iyer, P.}, \bibinfo{author}{Schoeny, A.},
  \bibinfo{author}{Sivakumar, K.}, \bibinfo{author}{Woodrich, D.},
  \bibinfo{author}{Yang, E.},  and \bibinfo{author}{Zhu, S.}
  (\textbf{\bibinfo{year}{2021}}). \enquote{\bibinfo{title}{Bounding the number
  of calling animals with passive acoustics and reliable locations}}
  \bibinfo{journal}{J. Acoust. Soc. Am} \textbf{150}(10.1121/10.0004994),
  \bibinfo{pages}{1496--1504}.

\bibitem[{Spiesberger and Fristrup(1990)}]{pass_loc}
\bibinfo{author}{Spiesberger, J.},  and \bibinfo{author}{Fristrup, K.}
  (\textbf{\bibinfo{year}{1990}}). \enquote{\bibinfo{title}{Passive
  localization of calling animals and sensing of their acoustic environment
  using acoustic tomography}} \bibinfo{journal}{The American Naturalist}
  \textbf{135}, \bibinfo{pages}{107--153}.

\bibitem[{Spiesberger and Wahlberg(2002)}]{prob_dens}
\bibinfo{author}{Spiesberger, J.},  and \bibinfo{author}{Wahlberg, M.}
  (\textbf{\bibinfo{year}{2002}}). \enquote{\bibinfo{title}{Probability density
  functions for hyperbolic and isodiachronic location}} \bibinfo{journal}{J.
  Acoust. Soc. Am.} \textbf{112}, \bibinfo{pages}{3046--3052}.

\bibitem[{Spiesberger(1999)}]{loc_animal}
\bibinfo{author}{Spiesberger, J.~L.} (\textbf{\bibinfo{year}{1999}}).
  \enquote{\bibinfo{title}{Locating animals from their sounds and tomography of
  the atmosphere: experimental demonstration}} \bibinfo{journal}{J. Acoust.
  Soc. Am.} \textbf{106}, \bibinfo{pages}{837--846}.

\bibitem[{Spiesberger(2001)}]{rec_loc_err}
\bibinfo{author}{Spiesberger, J.~L.} (\textbf{\bibinfo{year}{2001}}).
  \enquote{\bibinfo{title}{Hyperbolic location errors due to insufficient
  number of receivers}} \bibinfo{journal}{J. Acoust. Soc. Am.} \textbf{109},
  \bibinfo{pages}{3076--3079}.

\bibitem[{Spiesberger(2008)}]{patent_sbe2}
\bibinfo{author}{Spiesberger, J.~L.} (\textbf{\bibinfo{year}{2008}}).
  \plainquote{\bibinfo{title}{Estimation methods for wave speed}}
  \bibinfo{howpublished}{U.S. Patent No. 7,363,191}.

\bibitem[{Spiesberger(2011)}]{patent_sbe3}
\bibinfo{author}{Spiesberger, J.~L.} (\textbf{\bibinfo{year}{2011}}).
  \plainquote{\bibinfo{title}{Methods for estimating location using signal with
  varying signal speed}} \bibinfo{howpublished}{U.S. Patent No. 8,010,314}.

\bibitem[{Spiesberger(2012)}]{patent_sbe4}
\bibinfo{author}{Spiesberger, J.~L.} (\textbf{\bibinfo{year}{2012}}).
  \plainquote{\bibinfo{title}{Methods and apparatus for computer-estimating a
  function of a probability distribution of a variable}}
  \bibinfo{howpublished}{U.S. Patent No. 8,311,773}.

\bibitem[{Spiesberger(2014)}]{patent_sbe5}
\bibinfo{author}{Spiesberger, J.~L.} (\textbf{\bibinfo{year}{2014}}).
  \plainquote{\bibinfo{title}{Methods and computerized machine for sequential
  bound estimation of target parameters in time-series data}}
  \bibinfo{howpublished}{U. S. Patent No. 8639469}.

\bibitem[{Spiesberger(2017)}]{sbir_phase2}
\bibinfo{author}{Spiesberger, J.~L.} (\textbf{\bibinfo{year}{2017}}).
  \plainquote{\bibinfo{title}{Final report, target localization using
  muti-static sonar with drifting sonobuoys}} \bibinfo{howpublished}{Contract
  N68335-12-C-000211}.

\bibitem[{Spiesberger(2021)}]{patent_clock_sync2}
\bibinfo{author}{Spiesberger, J.~L.} (\textbf{\bibinfo{year}{2021}}).
  \plainquote{\bibinfo{title}{Estimation of clock synchronization errors using
  time difference of arrival}} \bibinfo{howpublished}{US patent 10915137}.

\bibitem[{Spiesberger(2023)}]{patent_clock_sync1}
\bibinfo{author}{Spiesberger, J.~L.} (\textbf{\bibinfo{year}{2023}}).
  \plainquote{\bibinfo{title}{Estimation of clock synchronization errors using
  time difference of arrival}} \bibinfo{howpublished}{CA 3107173}.

\bibitem[{Stanistreet \emph{et~al.}(2013)Stanistreet, Risch, and
  Parijs}]{pass_acous2}
\bibinfo{author}{Stanistreet, J.~E.}, \bibinfo{author}{Risch, D.},  and
  \bibinfo{author}{Parijs, S. M.~V.} (\textbf{\bibinfo{year}{2013}}).
  \enquote{\bibinfo{title}{Passive acoustic tracking of singing humpback whales
  (megaptera novaeangliae) on a northwest atlantic feeding ground}}
  \bibinfo{journal}{Plos One} \dodoi{10.1371/journal.pone.0061263}.

\bibitem[{Statek()}]{maru_crystal}
\bibinfo{author}{Statek}. \plainquote{\bibinfo{title}{Cs-1v-sm crystal
  oscillator}} .

\bibitem[{Tyrell(1964)}]{tyrell}
\bibinfo{author}{Tyrell, W.~N.} (\textbf{\bibinfo{year}{1964}}).
  \emph{\bibinfo{title}{Design of acoustic systems}}
  (\bibinfo{publisher}{Pergammon}, \bibinfo{address}{Oxford}), pp.
  \bibinfo{pages}{65--86}.

\bibitem[{Urazghildiiev and Clark(2013)}]{comp_localiz}
\bibinfo{author}{Urazghildiiev, I.~R.},  and \bibinfo{author}{Clark, C.~W.}
  (\textbf{\bibinfo{year}{2013}}). \enquote{\bibinfo{title}{Comparative
  analysis of localization algorithms with application to passive acoustic
  monitoring}} \bibinfo{journal}{J. Acoust. Soc. Am} \textbf{134}(6),
  \dodoi{10.1121/1.4824683}.

\bibitem[{Warner \emph{et~al.}(2017)Warner, Dosso, and
  Hannay}]{warner_and_dosso}
\bibinfo{author}{Warner, G.~A.}, \bibinfo{author}{Dosso, S.~E.},  and
  \bibinfo{author}{Hannay, D.~E.} (\textbf{\bibinfo{year}{2017}}).
  \enquote{\bibinfo{title}{Bowhead whale localization using
  time-difference-of-arrival data from asynchronous recorders}}
  \bibinfo{journal}{J. Acoust. Soc. Am.} \textbf{141},
  \bibinfo{pages}{1921--1935}.

\bibitem[{Zimmer(2005)}]{zimmer_clicks}
\bibinfo{author}{Zimmer, W. M.~X., J. M. P. M. P. T. . T. P.~L.}
  (\textbf{\bibinfo{year}{2005}}). \enquote{\bibinfo{title}{Echolocation clicks
  of free-ranging cuvier's beaked whales (ziphius cavirostris)}}
  \bibinfo{journal}{J. Acoust. Soc. Am} \dodoi{10.1121/1.1910225}.

\bibitem[{Zimmer~WM(2005)}]{sperm_direction}
\bibinfo{author}{Zimmer~WM, Tyack~PL, J. M. M.~P.}
  (\textbf{\bibinfo{year}{2005}}). \enquote{\bibinfo{title}{Three-dimensional
  beam pattern of regular sperm whale clicks confirms bent-horn hypothesis}}
  \bibinfo{journal}{Journal of the Acoustical Society of America}
  \dodoi{10.1121/1.1828501}.

\end{thebibliography}

\end{document}